\documentclass[letterpaper,twocolumn,twoside]{IEEEtran}

\usepackage{multirow}
\usepackage{amsmath,graphicx}
\usepackage{latexsym,amssymb}
\usepackage{url}
\usepackage{relsize}
\usepackage{bbm}
\usepackage{cite}
\usepackage{algorithm}
\usepackage{algorithmicx}
\usepackage{algpseudocode}
\usepackage{adjustbox}

\newlength{\timelength}
\def\PoissonDist{\text{Poisson}}

\newlength{\mylength}
\title{Single-Photon Depth Imaging \\
Using a Union-of-Subspaces Model
\thanks{This material is based upon work supported
in part by a Samsung Scholarship, by the
US National Science Foundation under Grant No.\ 1422034, and
by the MIT Lincoln Laboratory Advanced Concepts Committee.}
\thanks{D. Shin and J. H. Shapiro are with the Department of Electrical
Engineering and Computer Science and the Research Laboratory of Electronics,
Massachusetts Institute of Technology, Cambridge, MA 02139 USA.}
\thanks{V. K. Goyal is with the Department of Electrical and Computer Engineering,
Boston University, Boston, MA 02215 USA.}
}

\author{Dongeek Shin, Jeffrey H. Shapiro, and Vivek K Goyal}

\begin{document}

\maketitle

\begin{abstract}
Light detection and ranging systems 
reconstruct scene depth from time-of-flight
measurements. 
For low light-level depth imaging applications, such
as remote sensing and robot vision, these
systems use single-photon detectors that resolve individual photon
arrivals. Even so, they must detect a large number of photons
to mitigate Poisson shot noise and reject anomalous photon detections
from background light.
We introduce a novel framework 
for accurate depth imaging
using a small number of detected photons
in the presence of an unknown amount of background light
that may vary spatially. 
It employs a Poisson observation model for the photon detections
plus a union-of-subspaces constraint on the
discrete-time flux
from the scene at any single pixel.
Together, they enable a greedy signal-pursuit algorithm to rapidly and
simultaneously converge on accurate estimates of scene depth
and background flux, without any assumptions on spatial
correlations of the depth or background flux.
Using experimental single-photon data, 
we demonstrate that
our proposed framework recovers 
depth features with 1.7 cm absolute error,
using 15 photons per image pixel
and an illumination pulse
with 6.7-cm scaled root-mean-square length.
We also show that our framework outperforms
the conventional pixelwise log-matched filtering,
which is a computationally-efficient approximation
to the maximum-likelihood solution,
by a factor of 6.1 in absolute depth error.
\end{abstract}


\section{Introduction}

A conventional
light detection and ranging (LIDAR) system,
which uses a pulsed light source and
a single-photon detector,
forms a depth image pixelwise
using the histograms of photon detection times. 
The acquisition times for such systems
are made long enough to detect
hundreds of photons per pixel
for the finely binned
histograms these systems require to do accurate depth estimation.
In this letter, we introduce a framework for
accurate depth imaging using only a small number
of photon detections per pixel,
despite the presence of an unknown amount of spatially-varying background light in the scene.
Our framework uses a Poisson observation model for the photon detections 
plus a union-of-subspaces constraint on the
scene's discrete-time flux at any single pixel.
Using a greedy signal-pursuit algorithm---a modification of
CoSaMP~\cite{needell2009cosamp}---we solve for accurate estimates of scene
depth and background flux.
Our method forms estimates pixelwise and thus avoids assumptions
on transverse spatial correlations that may hinder the ability to resolve
very small features.
Using experimental single-photon
data, 
we demonstrate that our proposed depth imaging framework
outperforms log-matched filtering, 
which is the maximum-likelihood (ML) depth estimator
given zero background light.

Because our proposed framework is photon-efficient
while using an estimator that is pixelwise
and without background calibration,
it can be useful for dynamic low light-level imaging applications,
such as environmental mapping using unmanned aerial vehicles.

\begin{figure*}[t]
\centering
\hspace{-1cm}
\centerline{\includegraphics[scale=0.55]{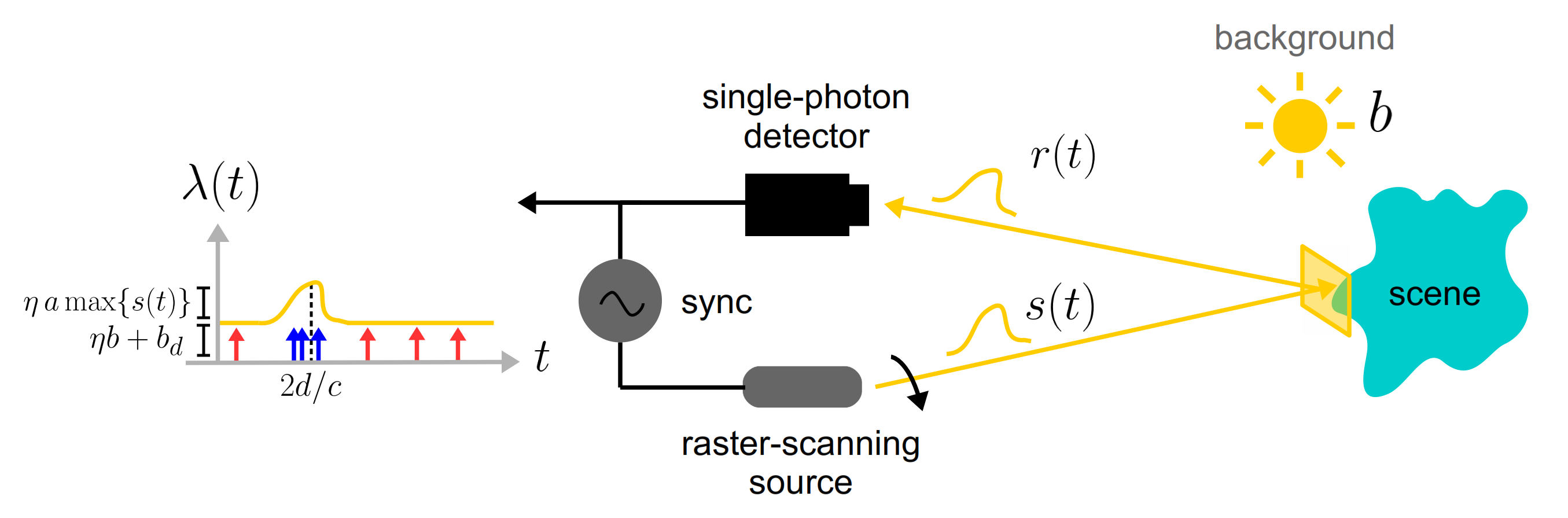}}
\caption{
An illustration of the single-photon imaging setup for one illumination pulse.
A pulsed optical source illuminates a scene pixel with photon-flux waveform $s(t)$.
The flux waveform $r(t)$ that is incident on the detector consists of the pixel return $as(t-2d/c)$---where $a$ is the pixel reflectivity, $d$ is the pixel depth, and $c$ is light speed----plus the background-light flux $b$.
The rate function $\lambda(t)$
driving the photodetection process equals
the sum of the pixel return and background flux,
scaled by the detector efficiency $\eta$, plus
the detector's dark-count rate $b_d$.
The record of detection times 
from the pixel return (or background light plus dark counts) is
shown as blue (or red) spikes,
generated by the Poisson process driven by $\lambda(t)$.
}
\label{fig_setup}
\vspace{-2mm}
\end{figure*}

\subsection{Prior Art}

The conventional LIDAR technique of estimating depth
using histograms of photon detections
is accurate when the number of photon detections is high.
In the low photon-count regime,
the depth solution is noisy due to shot noise.
It has been shown that
image denoising methods,
such as wavelet thresholding,
can boost the performance of 
scene depth recovery
in the presence of background noise~\cite{sun2005lidar}.
Also, using an imaging model that
incorporates occlusion constraints
was proposed to recover an accurate depth map~\cite{boufounos2012depth}.
However,
these denoising algorithms implicitly
assume that the observations are Gaussian distributed.
Thus, 
at low photon-counts,
where depth estimates are
highly non-Gaussian~\cite{mccarthy2009long}, their performance
degrades significantly~\cite{ShinKGS:2015}.

First-photon imaging (FPI)~\cite{kirmani2014first} is a framework
that allows high-accuracy imaging 
using only the first detected photon at every pixel.
It demonstrated that centimeter-accurate depth recovery
is possible by combining the non-Gaussian statistics of first-photon detection
with spatial correlations of natural scenes.
The FPI framework uses an imaging setup 
that includes a raster-scanning light source 
and a lensless single-photon detector.
More recently, 
photon-efficient
imaging frameworks that
use a detector array setup,
in which every pixel has the same acquisition time,
have also been proposed~\cite{ShinKGS:2015,shin2014icip,altmann2015lidar}.

We observe two common limitations 
that exist in the prior active imaging frameworks 
for depth reconstruction.
\begin{itemize}
\item 
\textbf{Over-smoothing: }
Many of the frameworks assume 
spatial smoothness of the scene
to mitigate the effect of shot noise.
In some imaging applications, however, 
it is important to capture fine spatial features 
that only occupy a few image pixels.
Using methods that assume spatial correlations 
may lead to erroneously over-smoothed images 
that wash out the scene's fine-scale features.
In such scenarios, a robust \textit{pixelwise} imager is preferable.
\vspace{2mm}
\item
\textbf{Calibration: }
Many imaging methods assume a 
calibration step to measure the amount of background flux
existing in the environment.
This calibration mitigates bias in
the depth estimate caused
by background-photon or dark-count detections, 
which have high temporal variance.
In practical imaging scenarios, however,
the background response varies in time,
and continuous calibration may not be
practical.
Furthermore, many methods assume background flux does not
vary spatially.
Thus,
a \textit{calibrationless} imager
that performs simultanous estimation of scene parameters
and spatially-varying background flux from photon detections is useful.
\end{itemize}

In this letter, we propose
a novel framework for depth acquisition
that is applied pixelwise
and without calibration.
At each pixel, our imager estimates 
the background response
along with scene depth from photon detections.
Similar to \cite{boufounos2012depth},
we use a union-of-subspaces constraint
for modeling the scene parameters.
However,
our union-of-subspace constraint is defined
for both the incoherent signal and background waveform parameters
that generate photon detections;
the constraint in \cite{boufounos2012depth}
is defined for only the coherent signal waveform
that is perturbed by Gaussian noise, not photon noise.

Using the derived imaging model,
we propose a greedy signal pursuit algorithm 
that accurately solves for the scene parameters at each pixel.
We evaluate the photon efficiency of this framework
using experimental single-photon data.
In the presence of strong background light,
we show that our pixelwise imager
gives an absolute depth error
that is $6.1$ times lower than that
of the pixelwise log-matched filter.

\section{Single-Photon Imaging Setup}

Figure.~\ref{fig_setup}
illustrates our imaging setup, for one illumination pulse,
in which photon detections are made.
A focused optical source,
such as a laser, 
illuminates a pixel of the scene
with the pulse waveform $s(t)$
that starts at time $0$ and has root-mean-square pulsewidth $T_p$.
This illumination is repeated
every $T_r$ seconds for a sequence of $N_s$ pulses.  
The single-photon detector, in conjunction with a time correlator,
is used to time stamp individual photon detections, relative to the time at which the immediately preceding pulse was transmitted.
These detection times,
which are observations of a time-inhomogeneous Poisson process,
whose rate function combines contributions from pixel return,
background light, and dark counts,
are used to estimate scene depth for the illuminated pixel.
This pixelwise acquisition process 
is repeated for $N_x\times N_y$ image pixels
by raster scanning the light source in the transverse directions.

\section{Forward Imaging Model}

In this section, we study the relationship
between the photon detections
and the scene parameters.
For simplicity of exposition and notation,
we focus on one pixel;
this is repeated for each pixel
of a raster-scanning or array-detection setup.

Let $a$, $d$, and $b$
be unknown scalar values that represent
reflectivity, depth,
and background flux at the given pixel.
The reflectivity value includes the effects of radial fall-off, view angle,
and material properties.
Then, after illuminating the scene pixel with
a single pulse $s(t)$,
the backreflected waveform that
is incident at the single-photon detector is 
\begin{align}
r(t)
= 
a s(t - 2 d /c) + b,
\hspace{\timelength}  
t\in [0,T_r).
\label{eq_reflected}
\end{align}

\subsection{Photodetection statistics}

Using (\ref{eq_reflected}),
we observe that the rate function that generates
the photon detections is
\begin{align}
\lambda(t)
= 
\eta
\left(
a s(t - 2 d /c) + b
\right) 
+ b_d,
\hspace{\timelength}  
t\in [0,T_r),
\end{align}
where $\eta \in (0,1]$ is the quantum efficiency
of the detector
and $b_d \geq 0$ is the dark-count rate 
of the single-photon detector.

Let $\Delta$ be the time bin duration
of the single-photon detector. Then,
we define $M=T_r/\Delta$ to be the total number of time bins
that capture photon detections.
Let $\mathbf{y}$ be the
vector of size $M \times 1$
that contains the number of photon detections at each time bin
after we illuminate the pixel $N_s$ times with pulse waveform $s(t)$.
Then, from photodetection theory~\cite{snyder1975random},
we can derive 
\begin{align}
\mathbf{y}_{k}
\sim
\PoissonDist
\!\left(
N_s 
\int_{(k-1)\Delta}^{k\Delta}
[\eta(
a s(t - 2d/c) 
+ b )+ b_d]
\
dt
\right),
\label{eq_photon}
\end{align}
for $k  = 1,\ldots,M$.
Note that we have assumed that our
total pixelwise acquisition time $N_sT_r$ is short enough
that $b$ is constant during that time.
Let
\begin{align}
\mathbf{v}_j
&=
\int_{(j-1)\epsilon}^{j\epsilon}\,
a \delta(t-2d/c)\, 
dt,
\label{approx_v}
\\
\mathbf{S}_{i,j}
&= 
\int_{(i-1)\Delta}^{i\Delta}\,
\int_{(j-1) \epsilon}^{j \epsilon}\,
N_s \eta {s}(t-y)
\,dt
\,dy,
\\
B
&=
\vphantom{\int_{(i-1)\Delta}^{i\Delta}}
N_s\Delta (\eta  b+b_d),
\end{align}
for $i=1,\ldots M, \ j=1,\ldots,N$,
where $\epsilon$ is a small number,
such that $T_r$ is divisible by $\epsilon$
and $N=T_r/\epsilon$. 
Defining $\mathbf{1}_{M\times1}$
to be an $M\times 1$ vector of ones,
we can approximate the rate function in (\ref{eq_photon})
and rewrite the distribution as
\begin{align}
\mathbf{y}_{k}
\sim
\PoissonDist
\big( \left(
\mathbf{S}\mathbf{v} + B\mathbf{1}_{M\times1}
\right)_k \big),
\label{obs_fin2}
\end{align}
for $k=1,\ldots M$. 
Finally, defining 
$\mathbf{A} = [\mathbf{S},\mathbf{1}_{M\times1}]$
and 
$\mathbf{x} = [\mathbf{v}^T,B]^T$,
we can further rewrite (\ref{obs_fin2}) as
\begin{align}
\mathbf{y}_{k}
\sim
\PoissonDist
\big( \left( \mathbf{A}\mathbf{x} \right)_k \big).
\label{obs_fin}
\end{align}
So far, we have simplified 
the pixelwise single-photon observation model,
such that the 
photon-count vector $\mathbf{y} \in \mathbb{N}^{M\times 1}$
is a linear measurement 
of scene response vector $\mathbf{x} \in \mathbb{R}_+^{(N+1)\times 1}$
corrupted by Poisson noise.

\subsection{Scene parameter constraints}
Using the expression in (\ref{approx_v}),
we observe that
\begin{align}
\mathbf{v}_j
=&
\int_{(j-1)\epsilon}^{j\epsilon}
a \delta(t - 2 d /c) \, dt
\\
=&
a \mathbbm{1}_{\{x:(j-1)\epsilon\leq 2x/c < j\epsilon \}}(d),
\end{align}
for $j=1,\ldots,N$,
where $\mathbbm{1}_A(x)$ is an indicator function that
equals $1$ if $x\in A$ and $0$ otherwise.
In other words, vector $\mathbf{v}$
has exactly one nonzero element,
and the value and index of the nonzero element
represents the scene reflectivity and depth at an image pixel, respectively.

We defined
our $(N+1)\times 1$ signal $\mathbf{x}$ 
to be a concatenation
of $\mathbf{v}$, which is the scene response vector of size $N$,
and $B$, which is the scalar representing background flux. 
Since $\mathbf{v}$ has exactly one nonzero entry,
$\mathbf{x}$ lies in the
union of $N$ subspaces defined as
\begin{align}
\mathcal{S}_{N}
=
\mathlarger\bigcup_{k=1}^{N}
\
\left\lbrace
\mathbf{x} \in \mathbb{R}^{N+1}
 :
\mathbf{x}_{\{1,2,\ldots,N\} \backslash \{k\}} = 0
\right\rbrace,
\end{align}
where each subspace is of dimension $2$.

\section{Solving the Inverse Problem}

Using accurate photodetection statistics and scene constraints,
we have interpreted the problem of 
robust single-photon depth imaging
as a noisy linear inverse problem,
where the signal of interest $\mathbf{x}$
lies in the union of subspaces $\mathcal{S}_{N}$.
Using (\ref{obs_fin}),
the observed photon count histogram $\mathbf{y}$
has the probability mass function
\begin{align}
f_Y(\mathbf{y};\mathbf{A},\mathbf{x})
=
\mathlarger\prod_{k=1}^M \
\dfrac{
e^{-(\mathbf{A}\mathbf{x})_k}
(\mathbf{A}\mathbf{x})_k^{\mathbf{y}_k}
}
{\mathbf{y}_k !}.
\end{align}
Thus, neglecting terms in the negative log-likelihood function
that are dependent on $\mathbf{y}$ but not on $\mathbf{x}$,
we can define the objective function
\begin{align}
\mathcal{L}(\mathbf{x};\mathbf{A},\mathbf{y})
= \mathlarger\sum_{k=1}^M  \,
\left[ 
\left( \mathbf{A}\mathbf{x} \right)_k
-\mathbf{y}_k \log \left(\mathbf{A}\mathbf{x} \right)_k
\right].
\end{align}
This objective can be proved to be convex in $\mathbf{x}$.

We solve for $\mathbf{x}$ 
by minimizing $\mathcal{L}(\mathbf{x};\mathbf{A},\mathbf{y})$
with the constraint that $\mathbf{x}$ lies in the
union of subspaces $\mathcal{S}_N$.
Also, because photon flux is a non-negative quantity,
the minimization results in a more accurate estimate
when we include a non-negative signal constraint.
In summary, the optimization problem 
that we want to solve
can be written as
\begin{align}
\label{opt}
\underset{\mathbf{x}}{\text{minimize }}\ \ &
\mathcal{L}(\mathbf{x};\mathbf{A},\mathbf{y}) \\
\text{s.t. } \ \
& \mathbf{x} \in \mathcal{S}_{N},
\nonumber \\
& \mathbf{x}_i \geq 0, \hspace{0.3cm} i=1,\ldots,(N+1)
\nonumber.
\end{align}

To solve our constrained optimization problem,
we propose an algorithm
that is inspired by existing fast greedy algorithms for
sparse signal pursuit.
CoSaMP~\cite{needell2009cosamp} is a
greedy algorithm that finds a
$K$-sparse approximate solution to a linear inverse problem.
We modify the CoSaMP algorithm
so that we obtain for a solution constrained to
the union of subspaces $\mathcal{S}_N$,
instead of a $K$-sparse one.

\begin{algorithm}
\caption{Single-photon depth imaging
using a union-of-subspaces model}
\begin{algorithmic}
\item \textbf{Input:} $\mathbf{y}$, $\mathbf{A}$, $\delta$
\item \textbf{Output:} $\mathbf{x}^{(k)}$
\item Initialize 
$\mathbf{x}^{(0)} \leftarrow \vec{0}$,
$\mathbf{u} \leftarrow \mathbf{y}$,
$k\leftarrow 0$;
\Repeat 
\State 
$k \leftarrow k+1$;
\State 
$\mathbf{\hat x} \leftarrow \mathbf{A}^T \mathbf{u}$;
\State 
$\Omega \leftarrow 
\text{supp}((\mathbf{\hat x}_{1:N})_{[1]}) 
\cup \text{supp}(\mathbf{x}^{(k-1)}_{1:N}) 
\cup \{N+1\}$;
\State 
$\mathbf{b}|_\Omega \leftarrow \mathbf{A}^{\dagger}_\Omega \mathbf{y}$;
\hspace{0.3cm}
$\mathbf{b}|_{\Omega^c} \leftarrow \mathbf{0}$;
\State
$\mathbf{x}^{(k)}
\leftarrow 
\mathcal{T}_0
\left(
[(\mathbf{b}_{1:N})_{[1]}^T, \hspace{1mm} \mathbf{b}_{N+1}]^T
\right)$
\State
$\mathbf{u} \leftarrow \mathbf{y}-\mathbf{A}\mathbf{x}^{(k)}$
\Until{$\| \mathbf{x}^{(k-1)} - \mathbf{x}^{(k)} \|_2^2 < \delta$}
\end{algorithmic}
\label{algorithm1}
\end{algorithm}

\begin{figure*}[t]
\centering
\begin{tabular}{c@{}c@{}c@{}c@{}c@{}c@{}c@{}c@{}c@{}c}
\hspace{-0.3cm}
&
{\small (a) Photograph}
&
\hspace{\mylength}
{\small (b) Truth}
&
\hspace{\mylength}
{\small (c) Log-matched filter}
&
\hspace{\mylength}
{\small (d) Proposed}
&
\hspace{\mylength}
{}
&
\hspace{\mylength}
{\small (e) Error of (c)}
&
\hspace{\mylength}
{\small (f) Error of (d)}
&
\hspace{\mylength}
{}
\vspace{0.1cm}
\\
\hspace{-0.3cm}
&
\includegraphics[scale=0.5]{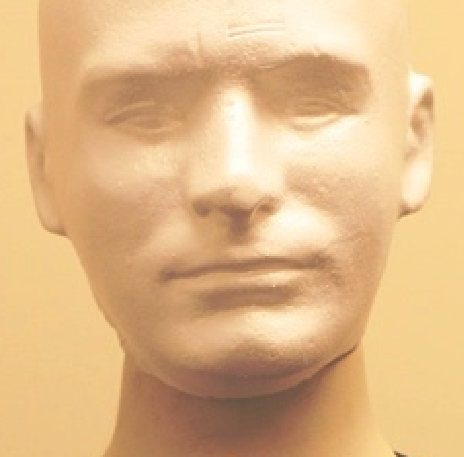} 
&
\hspace{\mylength}
\adjustbox{trim={0.22\width} {.12\height} {0.19\width} {.1\height},clip}%
{\includegraphics[scale=0.28]{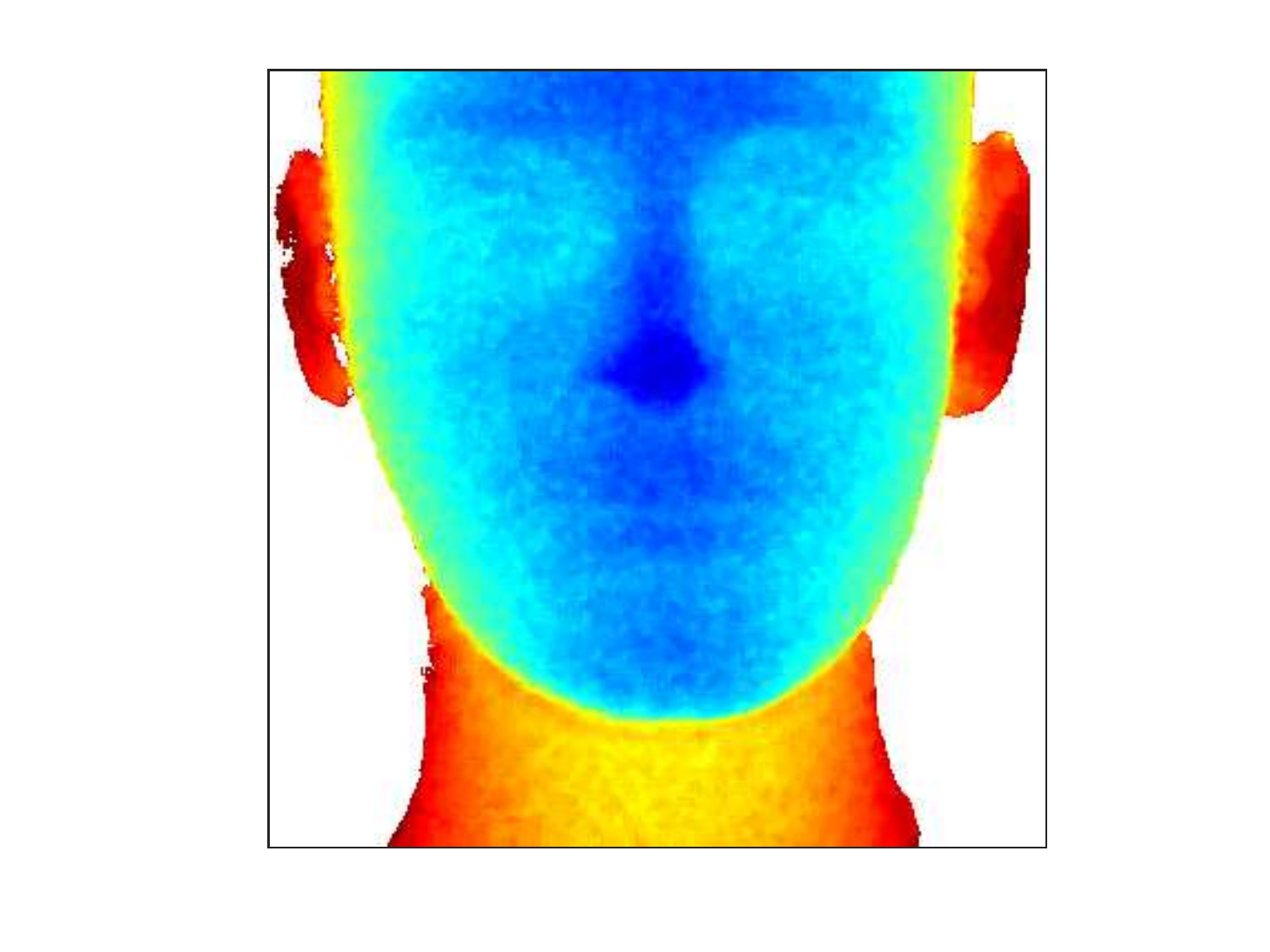}}
&
\hspace{\mylength}
\adjustbox{trim={0.22\width} {.12\height} {0.19\width} {.1\height},clip}%
{\includegraphics[scale=0.28]{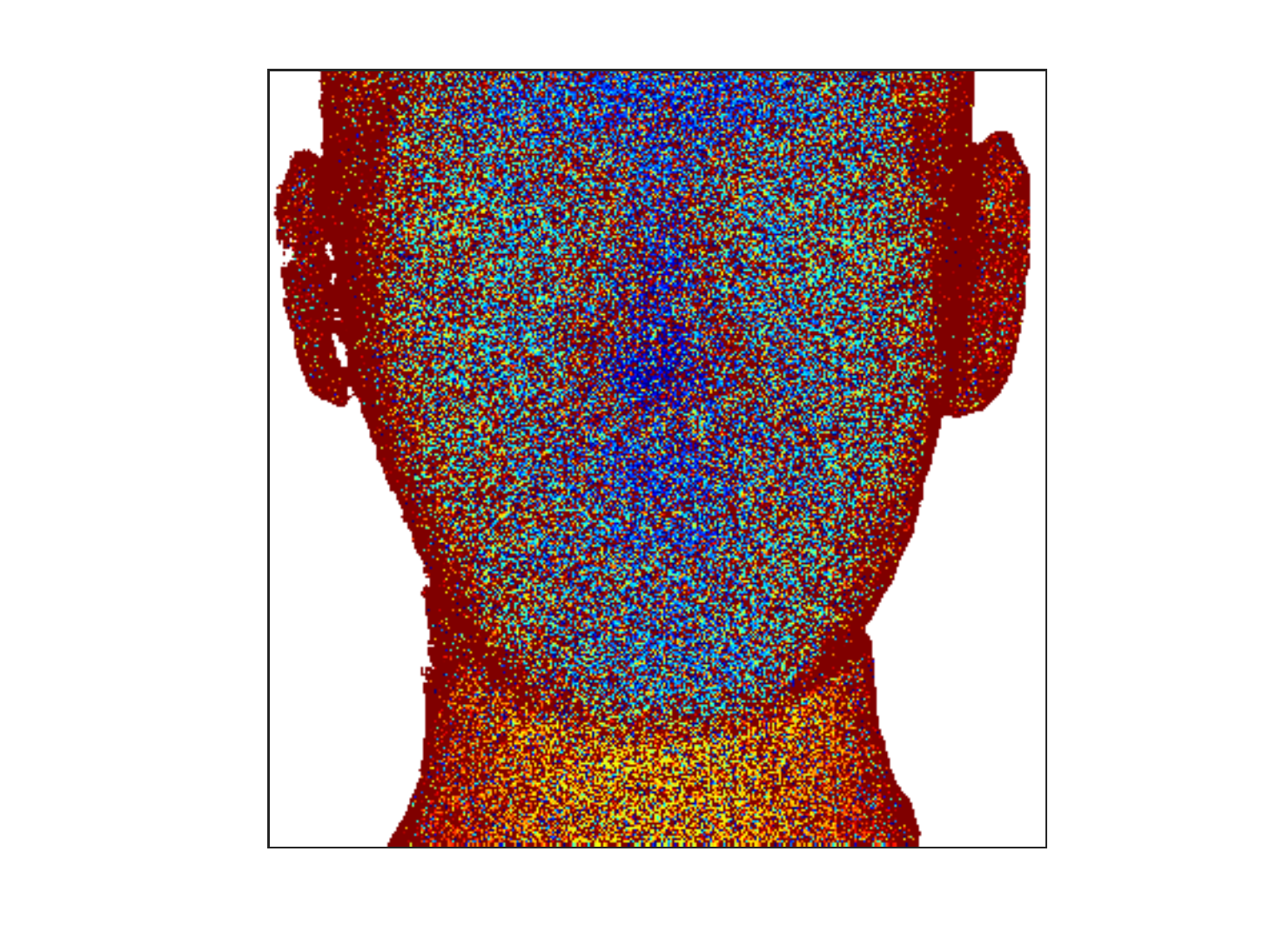}}
&
\hspace{\mylength}
\adjustbox{trim={0.22\width} {.12\height} {0.19\width} {.1\height},clip}%
{\includegraphics[scale=0.28]{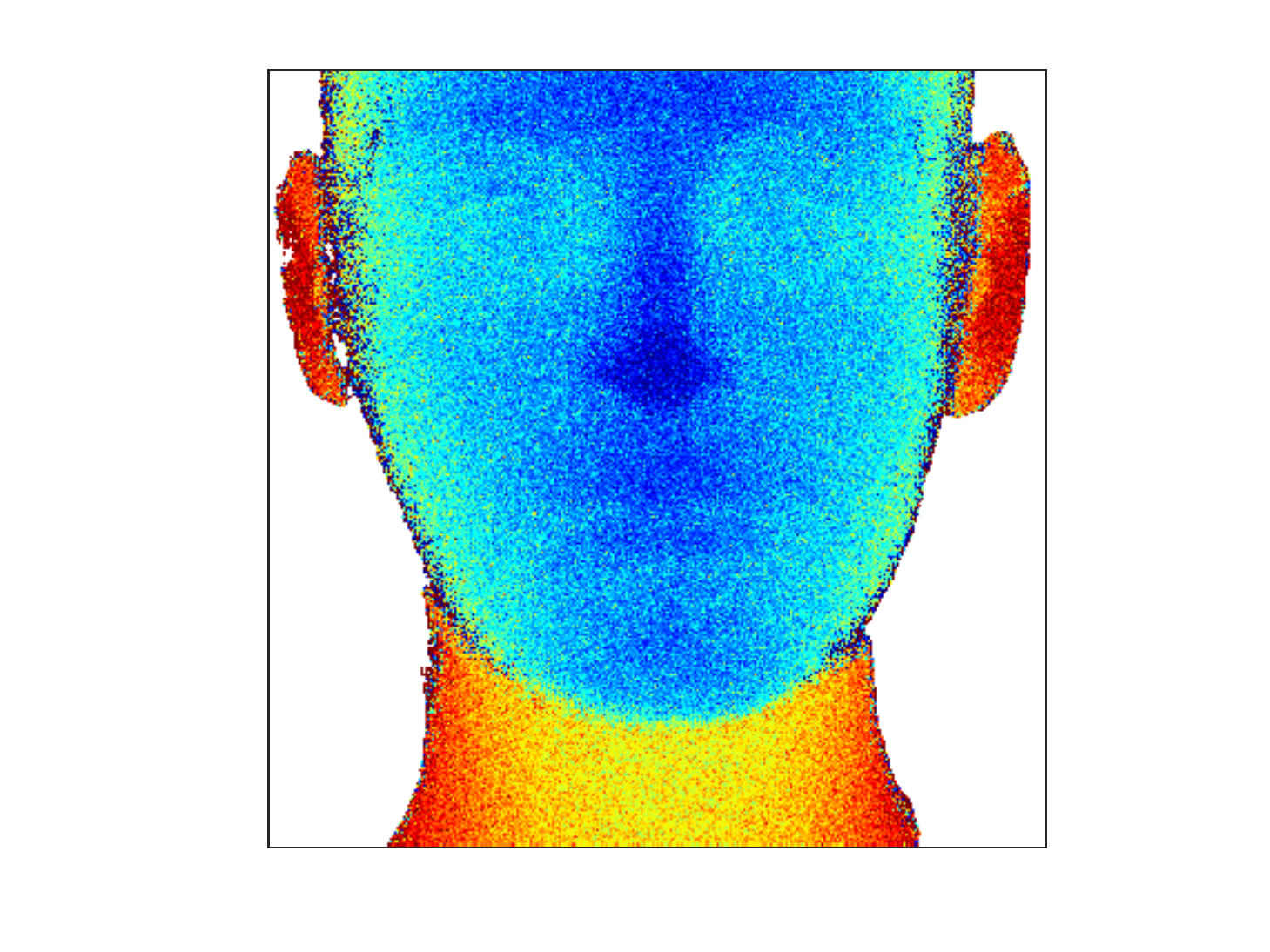}}
&
\hspace{\mylength}
\includegraphics[scale=0.26]{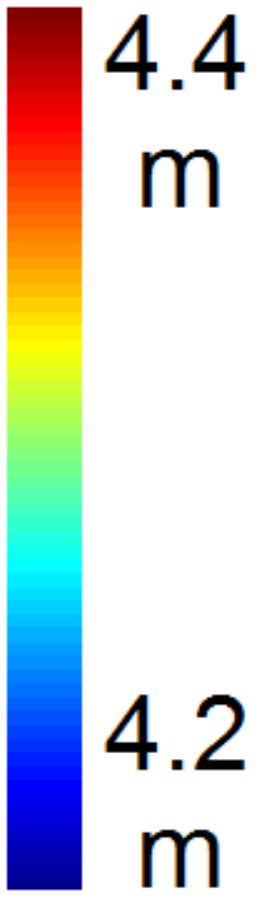}
&
\hspace{\mylength}
\adjustbox{trim={0.22\width} {.12\height} {0.19\width} {.1\height},clip}%
{\includegraphics[scale=0.28]{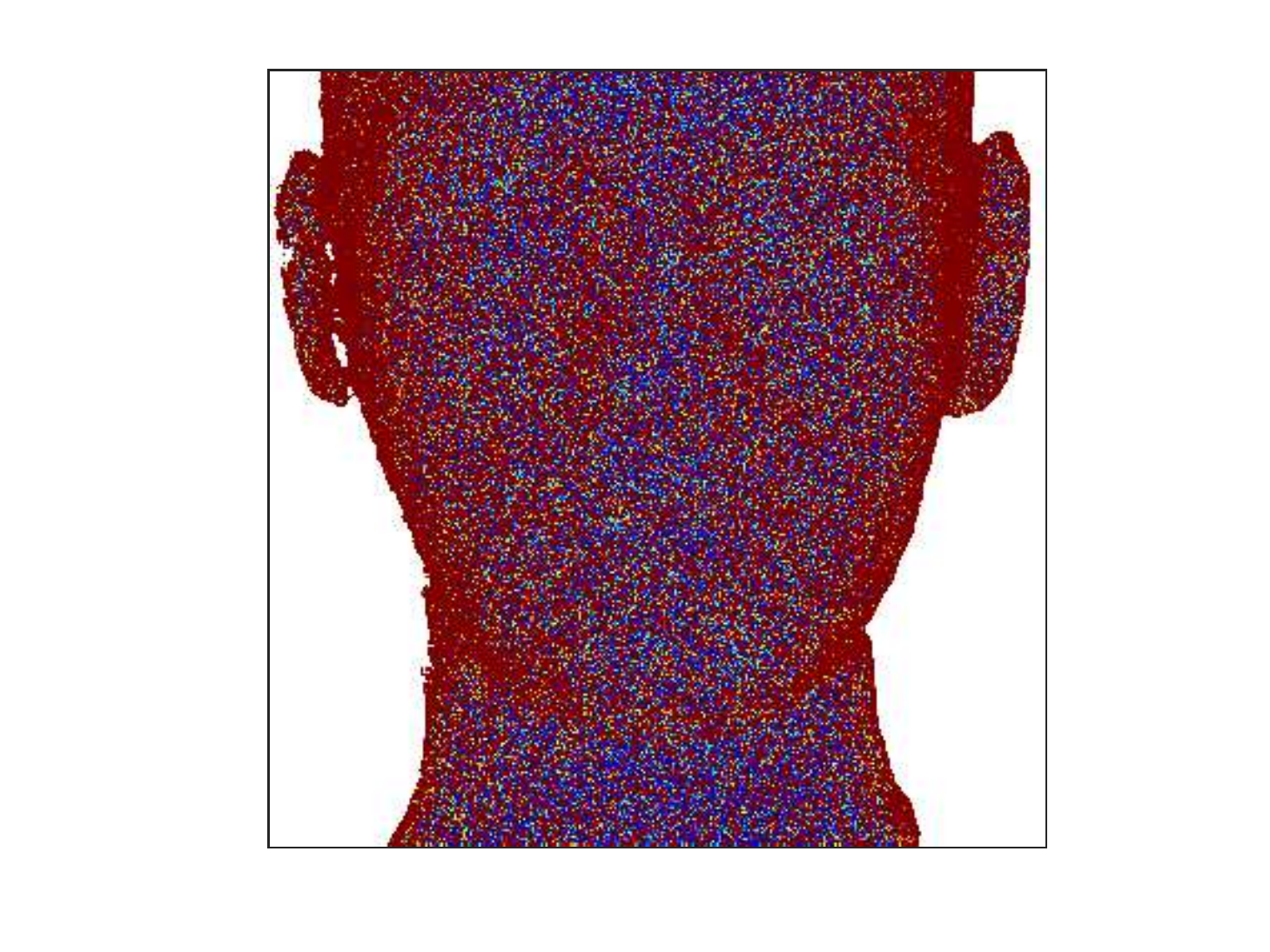}}
&
\hspace{\mylength}
\adjustbox{trim={0.22\width} {.12\height} {0.19\width} {.1\height},clip}%
{\includegraphics[scale=0.28]{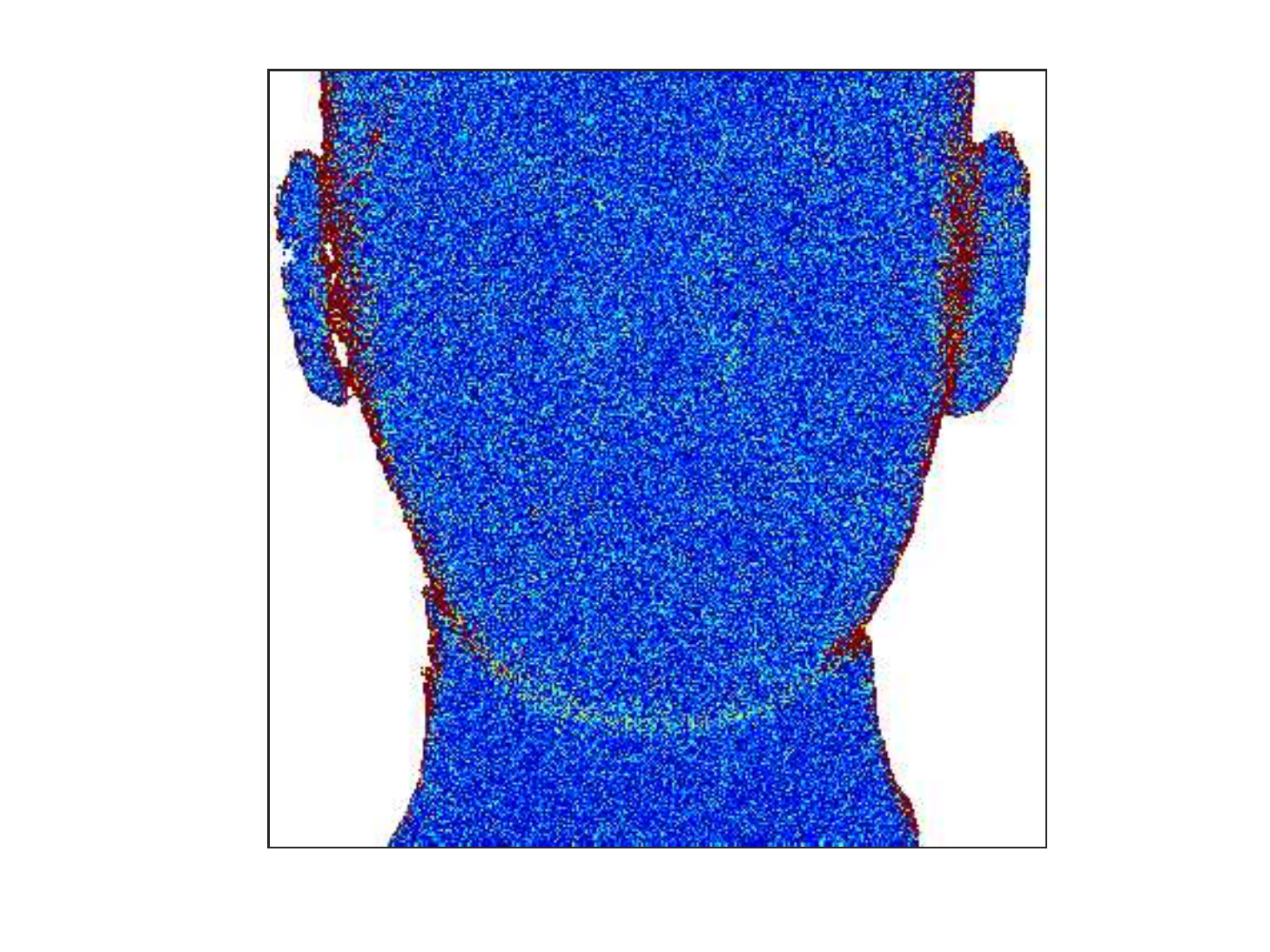}}
&
\hspace{\mylength}
\includegraphics[scale=0.26]{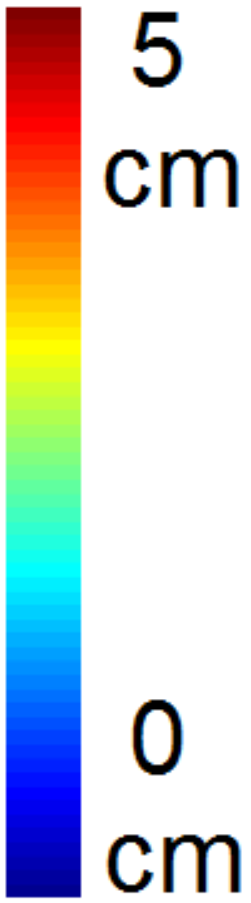}
\end{tabular}
\caption{
Experimental pixelwise depth imaging results using single photon observations.
The number of photon detections at every pixel was set to be $15$.
The figure shows the (a) photograph of imaged face,
(b) ground-truth depth,
(c) depth from log-matched filtering, which is approximately ML
and (d) depth using our method.
Also, (e) and (f) show the absolute depth error maps 
for ML and our framework, respectively.
}
\label{results1}
\vspace{-2mm}
\end{figure*}

Our proposed greedy algorithm is given in Algorithm~\ref{algorithm1}.
We define $\mathcal{T}_0(\mathbf{x})$ to be the thresholding operator
setting all negative entries of $\mathbf{x}$ to zero,
$\text{supp}(\mathbf{x})$ to be the 
set of indices of nonzero elements of $\mathbf{x}$,
and $\mathbf{x}_{[k]}$ to be the 
vector that approximates $\mathbf{x}$ with its $k$ largest terms.
Also, we take $\mathbf{A}_S$ to be a matrix with columns of $\mathbf{A}$
chosen by the index set $S$.
Finally, we use
$\mathbf{A}^T$ and 
$\mathbf{A}^\dagger$ 
to denote the
transpose and 
pseudo-inverse of matrix $\mathbf{A}$,
respectively.

To solve the constrained optimization problem in (\ref{opt}),
our algorithm iteratively performs
the following steps:
\begin{enumerate}
\item
gradient descent on
$\mathcal{L}(\mathbf{x};\mathbf{A},\mathbf{y})$,
which is approximated by the squared $\ell_2$-norm
$\|\mathbf{y}-\mathbf{A}\mathbf{x}\|_2^2$
for computational efficiency;
\item
projection of the intermediate estimate onto the closest subspace 
in the union of subspaces $\mathcal{S}_N$; and
\item
projection of the intermediate estimate onto the non-negative cone,
\end{enumerate}
until a convergence criterion is satisfied.
We define convergence of the solution as
$\| \mathbf{x}^{(k-1)} - \mathbf{x}^{(k)} \|_2^2 < \delta$,
where $\delta$ is a small number.

\section{Experimental results}

To validate our imaging framework, we
used a dataset collected by D. Venkatraman for
the First-Photon Imaging project~\cite{kirmani2014first};
this dataset and others are available from~\cite{github-fpi}.
The experimental setup uses a pulsed laser diode
with pulsewidth $T_p = 270 \text{ ps}$ and repetition period 
$T_r = 100 \text{ ns}$. 
A two-axis galvo was used to scan 
$350 \times 350$ pixels of a mannequin face
at a distance of about $4 \text{ m}$.
A lensless SPAD detector with quantum efficiency $\eta = 0.35$
was used for detection.
The background light level was set
using an incandescent lamp.
The original mannequin data from~\cite{github-fpi}
had the background count rate approximately equal
to the signal count rate. Our experiment uses
the cropped data showing only the mannequin's face,
where the background count rate
was approximately $0.1$ of the average signal count rate. 
Although we used a raster-scanning setup for our
single-photon imaging experiments,
since our imaging algorithm is applied pixelwise,
it can be also used for imaging with a floodlight illumination source 
and a single-photon detector array.
 
We could compare our imaging method
with the ML estimator of scene parameters.
Unfortunately, due to nonzero background flux,
ML estimation of $a$, $b$, and $d$ requires
minimizing a non-convex cost function,
leading to a solution without an accuracy guarantee.
Thus, zero background is assumed conventionally such that
the ML depth estimate reduces to 
the simple log-matched filter~\cite{blahut1987principles}:
\begin{align}
\label{logmatch}
\hat d_\text{ML}
= 
\dfrac{\epsilon}{2}
\left(
\underset{i\in\{1,\ldots,n\}}{\text{arg\,max}}
\log \mathbf{S}_i^T \mathbf{y}
\right).
\end{align}
We use (\ref{logmatch}) as the baseline depth estimator
that is compared with our proposed estimator.

Figure~\ref{results1} shows 
the results of recovering depth of the mannequin face
using single-photon observations.
The kernel matrix $\mathbf{S}$
was obtained by an offline measurement of the pulse shape.
Note that this measurement
depends only on the source, not on properties of the scene.
The ground-truth depth,
shown in Fig.~\ref{results1}(b),
was generated separately
by using background-calibrated ML estimation
from $200$ photons at each pixel.

In our depth imaging experiment,
the number of photon detections at each pixel was 
set to $15$.
We observe that,
due to extraneous background photon detections,
the log-matched filter estimate in Fig.~\ref{results1}(c)
($\text{average absolute error}=10.3 \text{ cm}$)
is corrupted with high-variance noise
and the facial features of the mannequin 
are not visible. On the other hand,
our estimate, shown in Fig.~\ref{results1}(d),
shows high-accuracy depth recovery
($\text{average absolute error}=1.7 \text{ cm}$).
As shown by the error maps in Fig.~\ref{results1}(e), (f),
both methods fail in depth recovery 
in the face boundary regions,
where very little light is reflected back from the scene
to the single-photon detector
and the signal-to-background ratio is thus very low. 
Also, we observe that
our estimated average background level over all pixels
was $\hat B = 1.4\times 10^{-3}$,
which is very close to the calibrated true background level
$B=1.3\times 10^{-3}$.
In this experiment, we had $M=N=801$.
Also, we set $\delta=10^{-4}$
and the average number of iterations until convergence
was measured to be $2.1$ over all pixels.
Code and data used to generate results
can be downloaded from \cite{github-uos}.

\section{Conclusions and Future Work}

In this letter,
we presented an imaging framework
for calibrationless, pixelwise depth reconstruction
using single-photon observations.
Our imaging model combined photon detection statistics
with the discrete-time flux constraints
expressed using a union-of-subspaces model.
Then, using our imaging model, 
we developed a greedy algorithm
that recovers scene depth
by solving a constrained optimization problem.

Our pixelwise imaging framework 
can be used in low light-level imaging applications, 
where the scene being imaged has
fine features and filtering techniques 
that exploit patchwise smoothness can 
potentially wash out those details.
For example, it can be useful in applications such as
airborne remote sensing~\cite{nilsson1996estimation},
where the aim is to recover finely-featured 3D terrain maps.

It is straightforward to generalize
the proposed single-photon imaging framework
to multiple-depth estimation,
where more than one reflector may be present at each pixel.
In the case of estimating depths of $K$ reflectors at a pixel, 
the $1$-sparsity assumption
must be changed to
more general $K$-sparsity assumption
when defining the union-of-subspaces constraint.


\begin{thebibliography}{11}

\bibitem{needell2009cosamp}
D.~Needell and J.~A. Tropp, ``{CoSaMP}: Iterative signal recovery from
  incomplete and inaccurate samples,'' \emph{Appl. Comput. Harmon. Anal.},
  vol.~26, no.~3, pp. 301--321, 2009.

\bibitem{sun2005lidar}
B.-Y. Sun, D.-S. Huang, and H.-T. Fang, ``Lidar signal denoising using
  least-squares support vector machine,'' \emph{IEEE Signal Process. Lett.},
  vol.~12, no.~2, pp. 101--104, 2005.

\bibitem{boufounos2012depth}
P.~T. Boufounos, ``Depth sensing using active coherent illumination,'' in
  \emph{Proc. IEEE Int. Conf. Acoust., Speech, and Signal Process.}, 2012, pp.
  5417--5420.

\bibitem{mccarthy2009long}
A.~McCarthy, R.~J. Collins, N.~J. Krichel, V.~Fern{\'a}ndez, A.~M. Wallace, and
  G.~S. Buller, ``Long-range time-of-flight scanning sensor based on high-speed
  time-correlated single-photon counting,'' \emph{Appl. Optics}, vol.~48,
  no.~32, pp. 6241--6251, 2009.

\bibitem{ShinKGS:2015}
D.~Shin, A.~Kirmani, V.~K. Goyal, and J.~H. Shapiro, ``Photon-efficient
  computational 3-d and reflectivity imaging with single-photon detectors,''
  \emph{IEEE Trans. Comput. Imaging}, vol.~1, 2015, to appear.

\bibitem{kirmani2014first}
A.~Kirmani, D.~Venkatraman, D.~Shin, A.~Cola{\c{c}}o, F.~N. Wong, J.~H.
  Shapiro, and V.~K. Goyal, ``First-photon imaging,'' \emph{Science}, vol. 343,
  no. 6166, pp. 58--61, 2014.

\bibitem{shin2014icip}
D.~Shin, A.~Kirmani, V.~K. Goyal, and J.~H. Shapiro, ``Computational 3d and
  reflectivity imaging with high photon efficiency,'' in \emph{Proc. IEEE Int.
  Conf. Image Process.}, Oct. 2014, pp. 46--50.

\bibitem{altmann2015lidar}
Y.~Altmann, X.~Ren, A.~McCarthy, G.~S. Buller, and S.~McLaughlin,
``Lidar waveform based analysis of depth images constructed using sparse single-photon data," {arXiv preprint arXiv:1507.02511}, 2015.

\bibitem{snyder1975random}
D.~L. Snyder, \emph{Random Point Processes}.\hskip 1em plus 0.5em minus
  0.4em\relax Wiley, New York, 1975.

\bibitem{github-fpi}
``{GitHub} repository for photon-efficient imaging,''
  \url{https://github.com/photon-efficient-imaging/sample-data/}.

\bibitem{blahut1987principles}
R.~E. Blahut, \emph{Principles and Practice of Information Theory}.\hskip 1em
  plus 0.5em minus 0.4em\relax Addison-Wesley Longman Publishing Co., Inc.,
  1987.

\bibitem{github-uos}
``{GitHub} repository for union-of-subspace imaging,''
  \url{https://github.com/photon-efficient-imaging/uos-imaging/}.

\bibitem{nilsson1996estimation}
M.~Nilsson, ``Estimation of tree heights and stand volume using an airborne
  lidar system,'' \emph{Remote Sensing of Environment}, vol.~56, no.~1, pp.
  1--7, 1996.


\end{thebibliography}
\end{document}